\documentclass[a4paper,10pt]{article}
\usepackage[utf8]{inputenc}
\newcommand{\be}{\begin{equation}}
\newcommand{\ee}{\end{equation}}
\usepackage{graphicx}
\usepackage{amssymb,amsmath,amsfonts}
\usepackage{color}

\def\<{\langle}
\def\>{\rangle}
\def\(({\left(}
\def\)){\right)}
\def\[[{\left[}
\def\]]{\right]}

\DeclareMathOperator*{\argmax}{argmax}

\title{Mismatching as a tool to enhance algorithmic performances of Monte Carlo methods for the planted clique model}
\author{Maria Chiara Angelini$^{1,2}$, Paolo Fachin$^1$, Simone de Feo$^1$}
\date{%
    $^1$ \footnotesize Dipartimento di Fisica, ``Sapienza, Universit\`{a} di Roma'', P.le A. Moro 5, 00185, Rome, Italy\\%
    $^2$ \footnotesize INFN, Sezione di Roma1, P.le A. Moro 5, 00185, Rome, Italy%
}

\begin{document}

\maketitle

\begin{abstract}
Over-parametrization was a crucial ingredient for recent developments in inference and machine-learning fields. However a good theory explaining this success is still lacking. 
In this paper we study a very simple case of mismatched over-parametrized algorithm applied to one of the most studied inference problem: the planted clique problem. We analyze a Monte Carlo (MC) algorithm in the same class of the famous Jerrum algorithm. We show how this MC algorithm is in general suboptimal for the recovery of the planted clique. We show however how to enhance its performances by adding a (mismatched) parameter: the temperature; we numerically find that this over-parametrized version of the algorithm can reach the supposed algorithmic threshold for the planted clique problem.
\end{abstract}

The deep learning revolution showed how over-parametrization can be a crucial ingredient to improve algorithmic performances in machine-learning, while a complete theory explaining such a success is still lacking \cite{goodfellow2016deep, carleo2019machine}.
Over-parametrized models are mismatched models, in the sense that they can express a different class of functions than the generative model that produced the data. While the analysis of general cases can be very involved, one could obtain very helpful insights from the study of simpler
models. In this paper we study a very simple case of mismatched over-parametrized algorithm:
We consider a standard inference problem for which we know the generative model.
In this case, as usually done in Bayesian inference, we can write the posterior probability distribution of the variables by using Bayes formula. This posterior probability can be interpreted as a Gibbs-Boltzmann (GB) measure over an ad-hoc Hamiltonian \cite{decelle2011PRE}. This is the Bayes optimal setting, where many results can be obtained using statistical physics tools \cite{zdeborova2016statistical}.
 A very simple way to introduce overparametrization is by adding an additional parameter, the temperature $T$. To recover the original Bayes posterior one could just take the value $T=1$, while studying the GB measure at temperatures different from 1 results in a mismatched model.
The introduction of the temperature modifies the free-energy landscape associated to the algorithms possibly changing their performances. 
Algorithms runned at $T\neq 1$ are in principle sub-optimal if one were able to sample exactly the GB measure. However sampling algorithms, taking place in a very high-dimensional space and typically running for a time much shorter than the equilibration time, rarely allow equilibrium sampling and their performances are not just related to the equilibrium free-energy. In these situations it is not obvious that their non-equilibrium sampling is optimal at $T=1$.
How do the performances of different algorithms change with the temperature $T$?
If Bayes optimal algorithms behave worse when $T$ is not equal to 1, we will show that other algorithms, such as Monte Carlo ones, become more efficient in the mismatched setting, as already suggested in ref. \cite{gamarnik2019landscape}.
We will show that entropic effects will help in finding the correct solution, showing that the only knowledge of the equilibrium free-energy is not enough to determine the performances of out-of-equilibrium algorithms.
In a statistical physics language, it is now well-known in optimization and inference problems that often the choice to work at $T>0$, that is to minimize the free-energy, leads to better results that working exactly at $T=0$, trying to directly minimize the energy. In this paper we show that finding the proper choice of the best $T$ is far from trivial.

In the following, we focus on a well known inference problem: the planted clique one. The planted clique problem is the following \cite{Jerrum92}:
Consider an Erdös–Renyi random graph of size $N$, in which each edge is present
independently with probability $1/2$. In such a random graph one plants a clique (a completely connected subgraph) of size $K$ manually adding any possible edge between two nodes belonging to the planted clique. Given a realization of such a graph, we want to identify
the planted clique. 
The identification of the planted clique is possible with an exhaustive search algorithm as long as $K>2\log_2(N)$, that is the size of the largest clique
present in a random graph \cite{grimmett1975colouring}. However many known algorithms are proved to fail in the regime $K/\sqrt{N}\to 0$,
including spectral algorithms \cite{alon1998finding}, Belief-Propagation based algorithms \cite{deshpande2015finding}, Sum of Squares Hierarchy \cite{barak2019nearly}:
the planted clique model displays one of the most well-studied computational-statistical
gaps in the literature of inference problems.

One could wonder how Monte Carlo (MC) algorithms work for this problem.
In ref. \cite{Jerrum92}, Jerrum introduced the problem, alongside with a MC algorithm  to find the planted clique. The allowed configuration space of Jerrum algorithm is the one containing perfect cliques of any size.
The author proves the failure of his MC algorithm for $K=o(\sqrt{N})$
but he does not prove its success for $K\ge\sqrt{N}$.
In ref. \cite{gamarnik2019landscape}, the authors analyze the performances of a different type of Monte-Carlo algorithm that spans the space of configurations at fixed magnetization $m=K$ (more details in the following sections),
including also "non-clique" configurations. They show that this type of MC
is not able to recover the planted clique below $K\le N^{2/3}$ 
in a polynomial time with $N$. However, they also show that the performance of the MC method can be salvaged all the way down to $K=\sqrt{N}$, using a mismatched magnetization $m>K$. 

One can also define a third MC method, in the same family of Jerrum one, that can be constructed directly from the Posterior distribution obtained from the Bayes formula: in the following we will call it BayesMC. The BayesMC was introduced in ref. \cite{angelini2018parallel}, together with an associated Parallel Tempering (PT) algorithm. The PT version seems to find the planted clique down to $K=2\log_2(N)$ in a polynomial time for relatively small $N$, but a theoretical analysis of it is still lacking.
In this work we will analyze the performances of the BayesMC: while at $T=1$ it is sub-optimal, increasing $T$ up to an optimal value, it can find the planted solution down to the regime $K=O(\sqrt{N})$. The results are in agreement with what conjectured in ref. \cite{gamarnik2019landscape}. Our reformulation of the mismatching in terms of a well defined statistical mechanics problem leads to an explicit physical intuition behind the newly introduced mismatched parameter, that in this case is just the temperature.

\section{The model and Monte Carlo algorithms}

\begin{figure}
\centering
\includegraphics[width=0.38\textwidth]{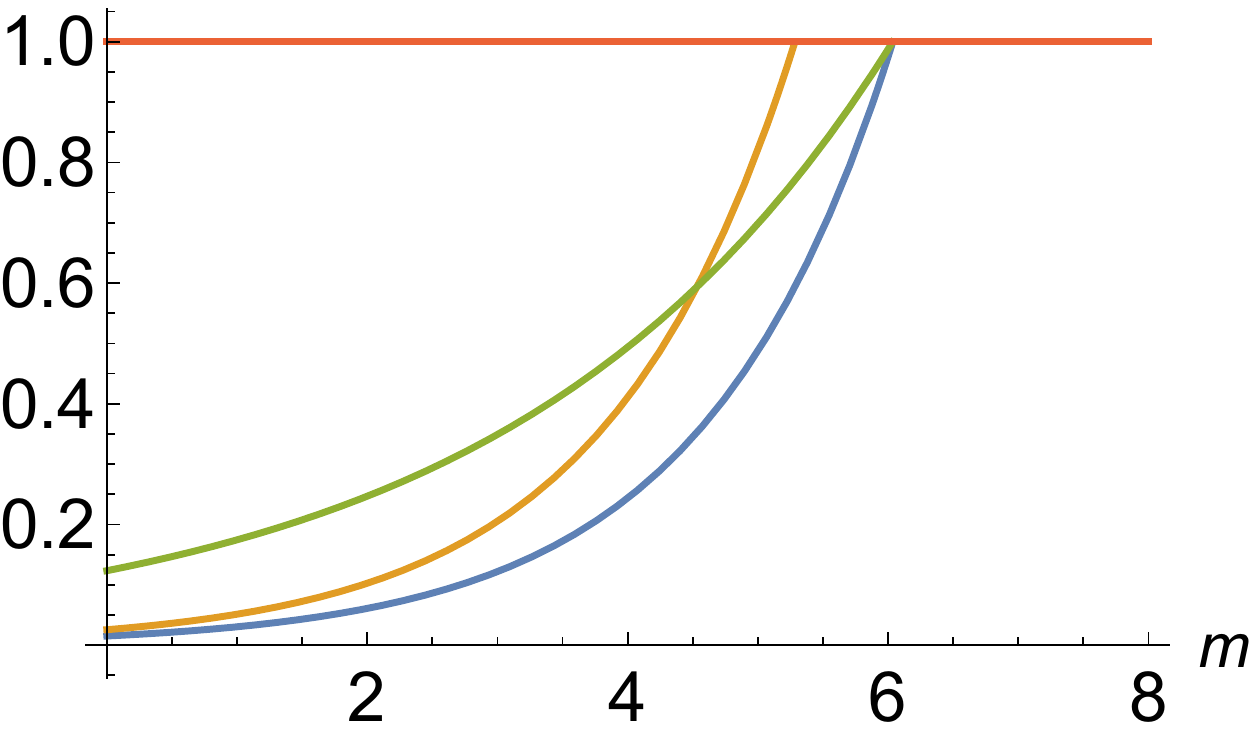}
\includegraphics[width=0.6\textwidth]{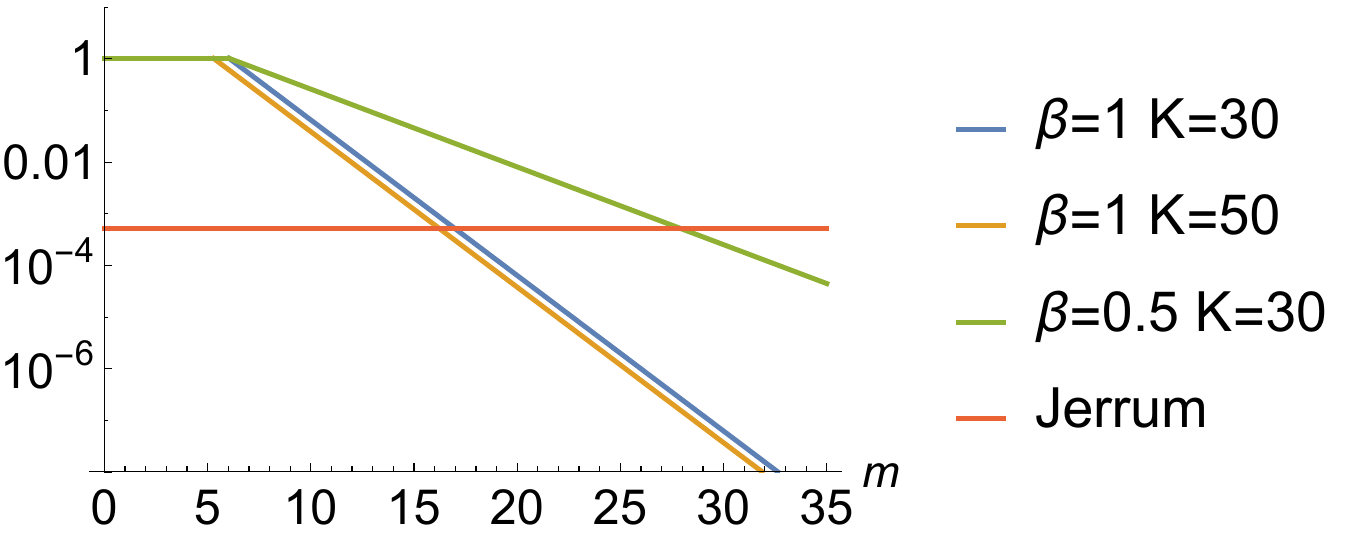}
\caption{\label{Fig:TransitionProb} Acceptance rate for the event $(x_i^n=0\to x_i^{n+1}=1)$ (left) and the event $(x_i^n=1\to x_i^{n+1}=0)$ (right) as a function of the actual magnetization $m$ at time $n$, for the BayesMC, defined in eqs. (\ref{eq:MC1},\ref{eq:MC2}) at different
values of $\beta$ and $K$, compared with the Jerrum algorithm defined in eq. (\ref{eq:Jerrum}), for a graph with $N=2000$. }
\end{figure}

To be concrete, we construct a graph of $N$ nodes with a planted clique $\mathcal{C}$ of size $K$. 
A label $v_i$ is associated with each node $i$: $v_i=1$ if node $i\in\mathcal{C}$, $v_i=0$ if node $i\notin\mathcal{C}$. 
We then extract the graph, univocally identified by its adjacency matrix  $\mathbf{A}$. 
Its element $A_{ij}$ takes the value 1 if an edge is put between node $i$ and node $j$, and the 
value 0 otherwise. The event $A_{ij}=1$ happens with the following probabilities:
\begin{equation}
p(A_{ij}=1|    \{{v\}})=
\begin{cases}
 1  &\text{if } v_iv_j=1 \\
 \frac{1}{2} &\text{otherwise}
\end{cases}.
\label{eq:likelihood}
\end{equation}
Given a realization of the graph, we want to estimate the labels $\{v\}$. 
We will perform this task using a MC algorithm, and we will indicate with $x_i=\{0,1\}$ the configuration if the $i$-th node explored by MC.
Following the Bayes formula, the posterior probability for $x_i$ given the graph is
\begin{equation}
P(x_i|\{{A\}})\propto P(A_{ij}|\{{x\}})P(x_i),
\label{eq:posterior}
\end{equation}
where the likelihood $P(A_{ij}|\{{x\}})$ is given by eq. (\ref{eq:likelihood}) and $P(x)$ is the \textit{prior} probability. 
The original problem has a global constraint on the size of the clique to be recovered (we are treating the case of known $K$):
$P(\{{x\}})$ should be zero if $\sum_i x_i\neq K$.
However, following Jerrum's idea, we want to construct a Monte-Carlo algorithm, for which the actual magnetization $m=\sum_i x_i$ is not fixed. 
We then replace the global constraint with a local prior on the single node\footnote{We are looking at the gran-canonical ensemble, in the language of statistical physics}: $P(x)=\left(\frac{K}{N}\right)^{x}\left(1-\frac{K}{N}\right)^{1-x}$.
At this point we introduce an additional parameter, the inverse temperature $\beta=\frac{1}{T}$,
and a new $\beta$-dependent distribution:
\begin{equation}
 P_{\beta}(\{x\}|\{{A\}})\propto P^\beta(\{x\}|\{{A\}})   
 \label{eq:betaPosterior}
\end{equation}
that in the case $\beta=1$ reduces to the standard posterior in  eq.(\ref{eq:posterior}).
We define an Hamiltonian associated to the posterior probability of eq. (\ref{eq:betaPosterior}) as: $P_{\beta}(\{x\}|\{{A\}})\equiv \frac{1}{\mathcal{N}}e^{-\beta H(\{{x\}})}$,
where $\mathcal N$ is a normalization factor and the Hamiltonian $H$ has the form:
\small
\begin{align}
\nonumber
H(\{{x\}})=&-\sum_i \log(P(x_i)) +\\
&-\sum_{ij}\left[\left(1-A_{ij}\right)\log\frac{\((1-x_ix_j\))}{2}+A_{ij}\log\frac{\((1+x_ix_j\))}{2}\right].
\label{eq:Hamiltonian}
\end{align}
\normalsize
Let us look carefully to the single terms in the Hamiltonian.
The first sum $\sum_i \log(P(x_i))$ acts as a local constant field on the single nodes.
Changes in the energy due to this term are of order $O(\sqrt{N}\log(\sqrt{N}))$ when $K=O(\sqrt{N})$, thus one could naively conclude that it is subdominant and it can be neglected. However we will show its effects in the MC acceptance rate, shown in Fig. \ref{Fig:TransitionProb}.  
The second term $-\sum_{ij}\left[\left(1-A_{ij}\right)\log\frac{\((1-x_ix_j\))}{2}\right]$
implies $H=\infty$ if there exists at least one couple $i,j$ for which $A_{ij}=0$ and $x_ix_j=1$, preventing from having configurations without links between two elements of a clique: it acts as a hard constraint on the allowed configurations. Finally the last term $-\sum_{ij}\left[A_{ij}\log\frac{\((1+x_ix_j\))}{2}\right]$ 
favours the addition of nodes to the clique whenever it is possible and it can be rewritten as $-\sum_{(i,j)\in E}\log(2)\left(x_ix_j-1\right)$, where $E$ is the set of edges of the graph.
Changes in the energy due to this term are extensive in the regime $K=O(\sqrt{N})$.
Summarizing, the Hamiltonian associated to the planted clique can be recast in the Hamiltonian of an Ising model with an external field plus an hard constraint on the allowed configurations.

One can thus define a standard Metropolis algorithm from the Hamiltonian in eq. (\ref{eq:Hamiltonian}), that we will call BayesMC, where each proposed move is accepted with probability
$\text{min}\left(1,e^{-\beta\Delta E}\right)$, with $\Delta E=H(\{{x^{n+1}\}})-H(\{{x^{n}\}})$ \cite{angelini2018parallel}. 
At time $n$ a spin $i$ is chosen at random among the $N$ ones. If its value is $x_i^n=0$, 
it is flipped with probability:
\begin{equation} 
 P(x_i^n=0\to x_i^{n+1}=1)=\begin{cases}
                             &0 \quad \text{ if } \exists j: x_j^n=1 \text{ and } A_{ij}=0\\
                             &\text{min}\left(e^{-\beta\[[\log(1-\frac{K}{N})-\log(\frac{K}{N})+m\log(\frac{1}{2})\]]},1\right) \text{ o.w.}
                           \end{cases},
\label{eq:MC1}                           
\end{equation}
$m=\sum_{j\neq i=1}^Nx_j^n$ being the actual magnetization.
On the other hand, if at time $n$ the value of the randomly chosen spin is $x_i^n=1$, 
it is flipped with probability:
\begin{equation}
P(x_i^n=1\to x_i^{n+1}=0)=\text{min}\left(e^{\beta\left[\log(1-\frac{K}{N})-\log(\frac{K}{N})+m\log(\frac{1}{2})\right]},1\right)
\label{eq:MC2}
\end{equation}
A single Monte-Carlo step (MCS) is the attempt to change $N$ spins.
The starting condition is the configuration $x_i^0=0 \quad \forall i$.
Being the system fully connected, the computation of the new energy after a flip of a spin
is of order $O(N)$. However, the proposal to flip a spin is accepted only $O(K)$
times: A single step of the algorithm thus takes $O(K \cdot N )$ time.
The algorithm stops when it finds a clique of size $K$: for the analyzed values of $K$, there are no random cliques of that size in the graph, thus one can be sure that when the algorithm stops it just found the planted clique. 

One could compare the BayesMC algorithm with the one introduced by Jerrum in ref. \cite{Jerrum92}, for which:
\begin{equation} 
 P_{\text{Jerrum}}(x_i^n=0\to x_i^{n+1}=1)=\begin{cases}
                             0 \quad &\text{ if } \exists j: x_j^n=1 \text{ and } A_{ij}=0\\
                             1 &\text{ otherwise}
                           \end{cases}
\label{eq:Jerrum}                           
\end{equation}
$$P_{\text{Jerrum}}(x_i^n=1\to x_i^{n+1}=0)=\lambda^{-1}$$
Jerrum suggests the choice $\lambda=N$ for which cliques of size greater
than $(2-\epsilon)\log_2(N)$ dominate when the process is in equilibrium: with this choice the equilibrium measure should be peaked on the planted clique.
While in the Jerrum MC the transition probabilities are fixed by the parameter $\lambda$, in the BayesMC the transition probabilities vary during the simulation 
depending on the actual magnetization $m$: a comparison between the acceptance rates for the two algorithms is shown in Fig. \ref{Fig:TransitionProb}. The more striking difference is the high acceptance rate for BayesMC w.r.t the Jerrum MC for the event $(x_i^n=1\to x_i^{n+1}=0)$ at small $m$, that allows to move fast between cliques of small sizes. 
\footnote{One could also choose not to follow detail balance, and to take a mix between Jerrum and BayesMC for the acceptance rates: just as an example, a particularly good choose could be Jerrum rate for the event $(x_i^n=0\to x_i^{n+1}=1)$ and BayesMC for the event $(x_i^n=1\to x_i^{n+1}=0)$. A similar way of reasoning is the one used to design heuristic algorithms widely used in satisfiability problems, like WALKSAT or ASAT \cite{PhysRevE.74.037702, seitz2005focused}. }

We will present a numerical study of the Jerrum algorithm in the appendix, while in the following sections we will focus our attention on the BayesMC, for which 
the physical meaning of the parameters is more explicit.

Let us just mention that one could also define a third MC method, that works at fixed magnetization $m\equiv K$. Defining an energy $E=\sum_{ij}(1-A_{ij})x_ix_j$ that associate a unitary cost to couples of unconnected nodes $i,j$ inserted in the putative planted clique, one can analyze the Metropolis algorithm associated to it at a given temperature (in this setting one should find the ground state to discover the planted clique). Working at fixed magnetization, the possible moves at time $n$ are the ones of the type $(x_a^n,x_b^n)=(1,0)\to (x_a^{n+1},x_b^{n+1})=(0,1)$. In ref. \cite{gamarnik2019landscape}, it is proved the impossibility for
 this class of MC algorithms to recover the planted clique for $K\le N^{2/3}$ 
in a polynomial time with $N$. However, the authors also show that the performance of this MC algorithm can be salvaged all the way down to $K=\sqrt{N}$, working with a mismatched fixed magnetization $m>K$. 

\section{Numerical Thresholds for the Monte Carlo algorithm}

\begin{figure}
\centering
\includegraphics[width=0.6\textwidth]{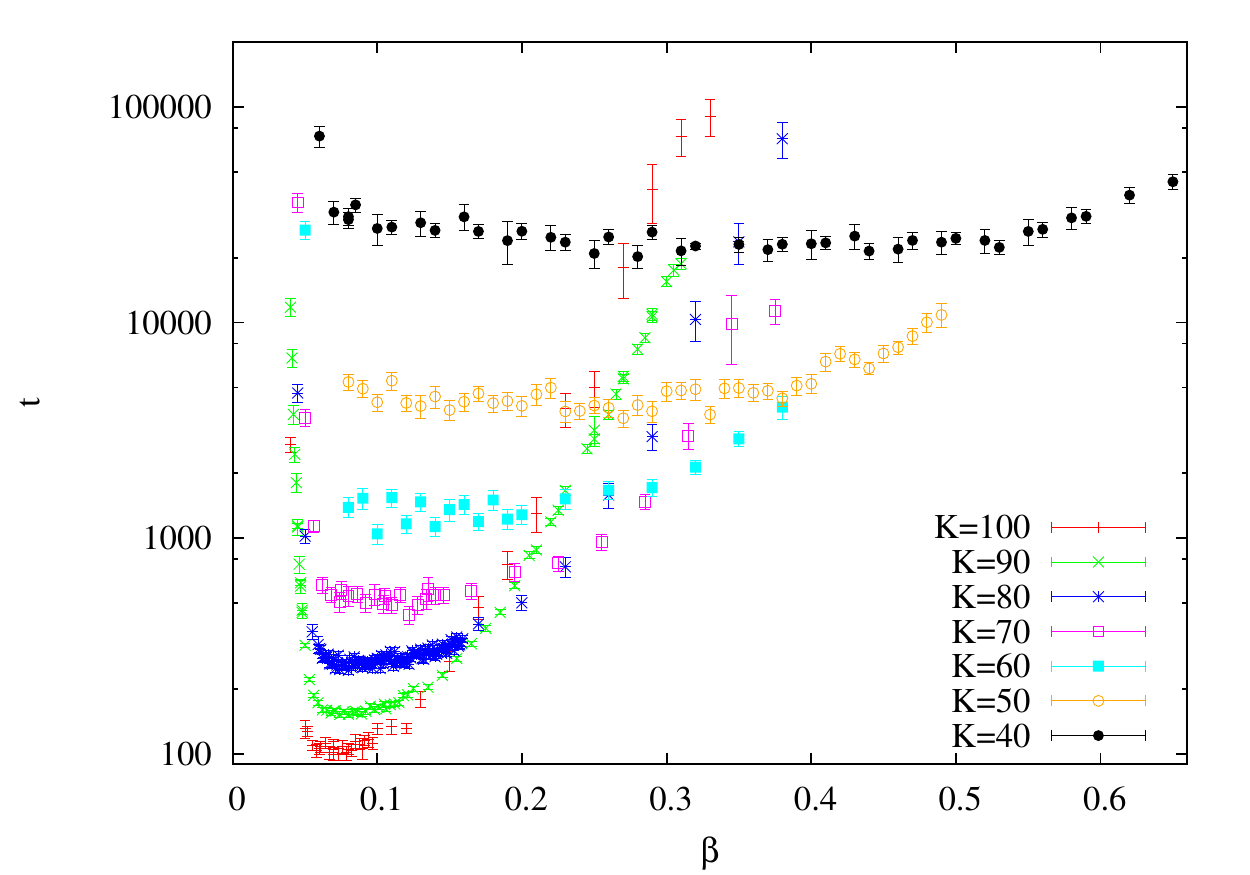}
\caption{\label{Fig:MC_beta} Average number of MCS $t$ needed to find the planted clique with BayesMC, as a function of $\beta$ for different $K$ at fixed $N=2000$. Averages are taken over $\sim100$ samples. For these values of $K$ and $N$, the planted clique is not recovered in $t\le10^7$ if $\beta=1$. }
\end{figure}

As already said, the BayesMC algorithm defined in eqs. (\ref{eq:MC1},\ref{eq:MC2})  that aims to sample the posterior distribution, should work at $\beta=1$. However, we numerically checked that for $\beta=1$ BayesMC does not succeed to find the planted clique in a reasonable time ($t<10^7$) even for $K>>\sqrt{N}$ at relatively small $N$ ($N\simeq 10^3$). For this reason, we skipped the analysis in this case and we moved directly to analyze the case of $\beta<1$. In Fig. \ref{Fig:MC_beta} we show $t$, that is the average number of MCS needed to find the planted clique, as a function of $\beta$ for different $K$ at fixed $N=2000$. 
\begin{figure}
\centering
\includegraphics[width=0.47\textwidth]{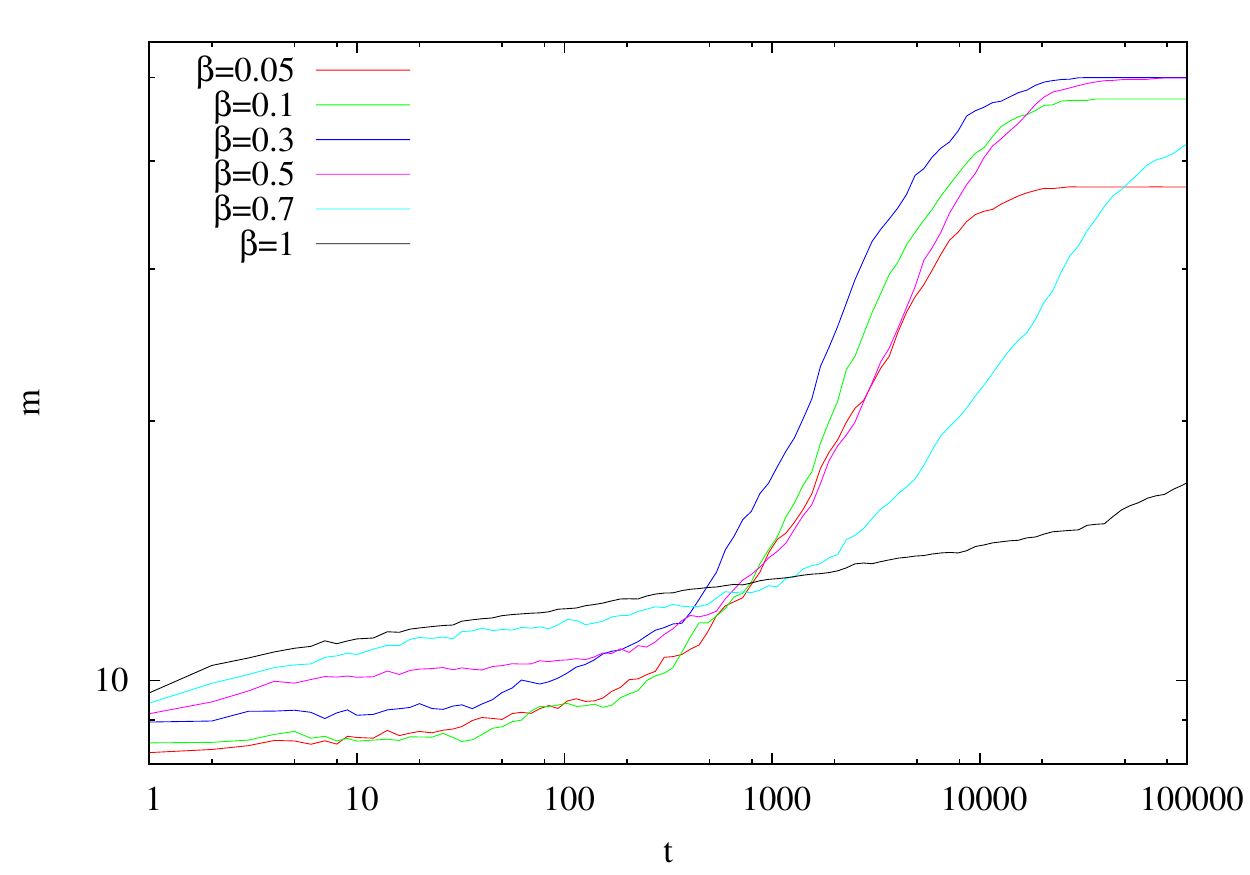}
\includegraphics[width=0.47\textwidth]{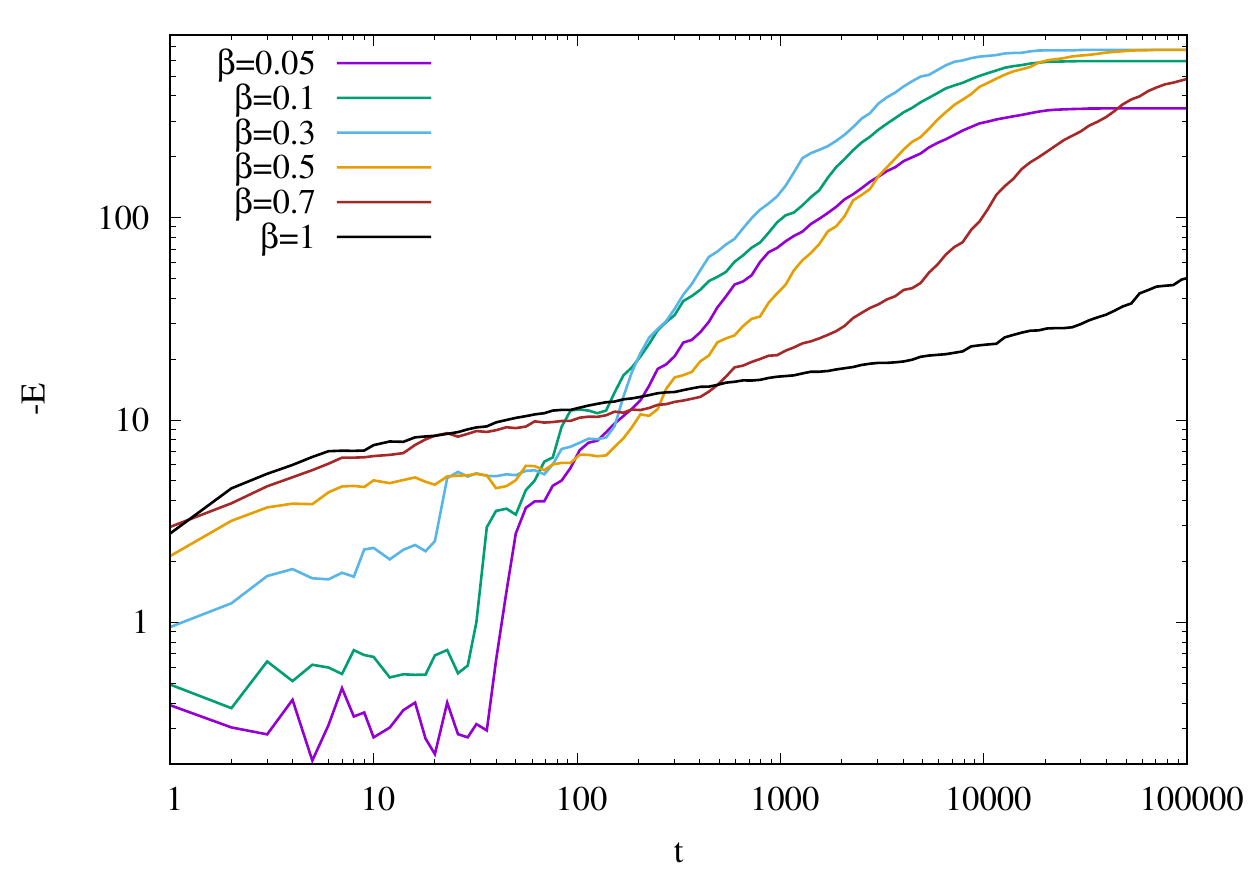}
\caption{\label{Fig:magn} Left: Average magnetization as a function of the number of MCS $t$ of BayesMC for samples with $N=2000$ and $K=50$, at different values of inverse temperature $\beta$.
In this case $\beta_{opt}\simeq 0.3$. Averages are taken over 200 samples.
Please note how for $T=1$, after $t=10^5$ steps, the average magnetization is still far from reaching $m=K$. 
Right: Average energy (with changed sign) for the same samples as in the Left Fig.}
\end{figure}
The minimum of $t$ is located at $\beta_{opt}(K)<1$: entropic effects seem to help speeding-up the convergence. 
For a deeper understanding of the function of $\beta$, in Fig. \ref{Fig:magn} we plot the average magnetization and energy as a function of $t$ at fixed value of $N$ and $K$ and for different inverse temperatures $\beta$. All the curves show the same qualitative behaviour: for small $t$, there is a first plateau at small magnetization $m_1\lesssim O(\log_2(N))$, of the order of the size of the most numerous random cliques. $m_1$ depends on $\beta$ and it is smaller for smaller $\beta$ (energy in the first plateau is larger for smaller $\beta$). We identify as $t^*(\beta)$ the time needed to escape this first plateau. If $\beta$ is too large, the time $t^*$ grows. In this case it is plausible that random cliques of larger sizes $m_1$ are trapping the dynamics, while working at smaller $\beta$ allows to explore smaller random cliques without large barriers between them. 
For $t>t^*(\beta)$, the magnetization grows until a second plateau at $m_2$ that corresponds to the equilibrium magnetization. For too small $\beta$, $m_2<K$, and the planted clique is found only as a rare fluctuation. Correspondingly, for too small $\beta$, the equilibrium energy is higher than that found at the optimal $\beta_{opt}(K)$, that corresponds to the planted ground state of the system.
The minimum $t$ needed to find the planted clique is obtained for a value  $\beta_{opt}(K)$ that is a compromise between these two opposite effects.

\begin{figure}
\centering
\includegraphics[width=0.48\textwidth]{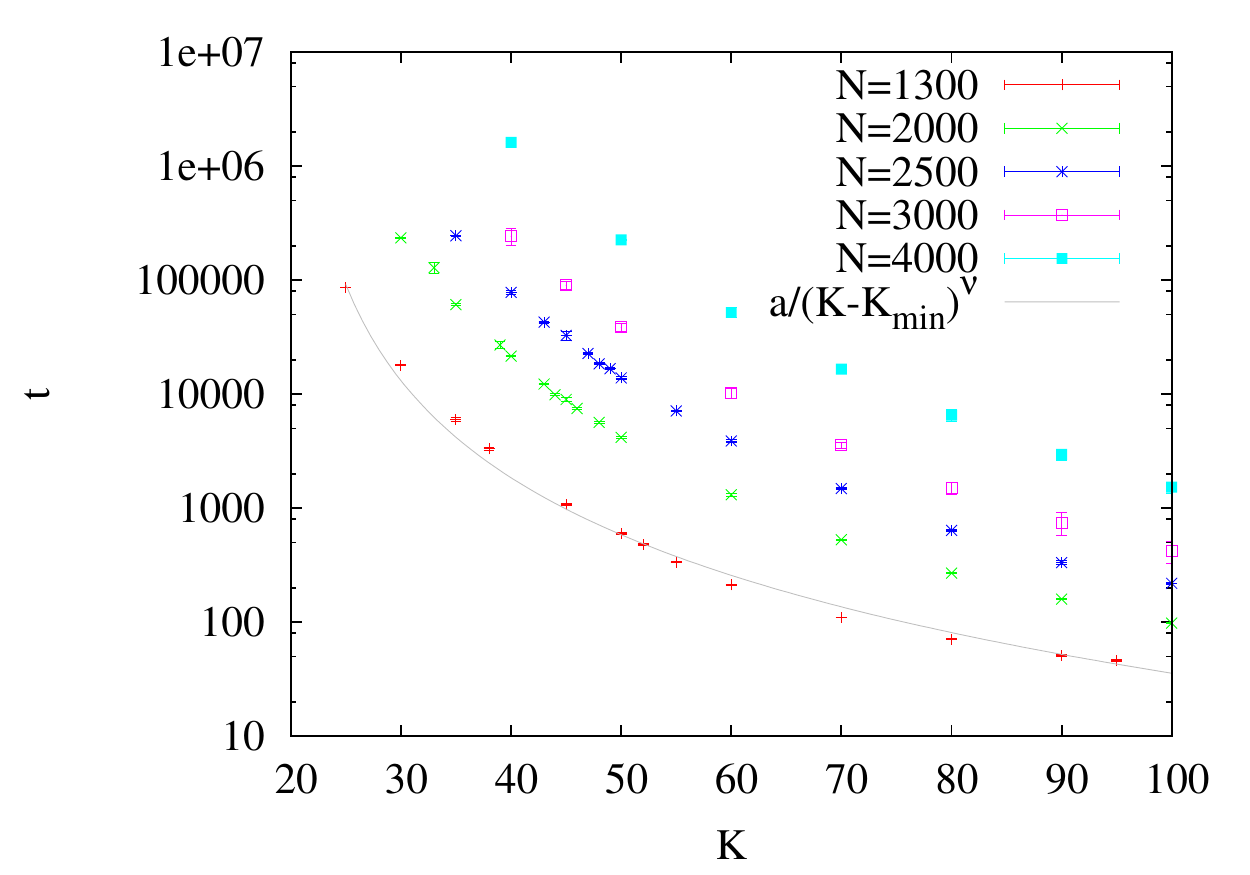}
\includegraphics[width=0.48\textwidth]{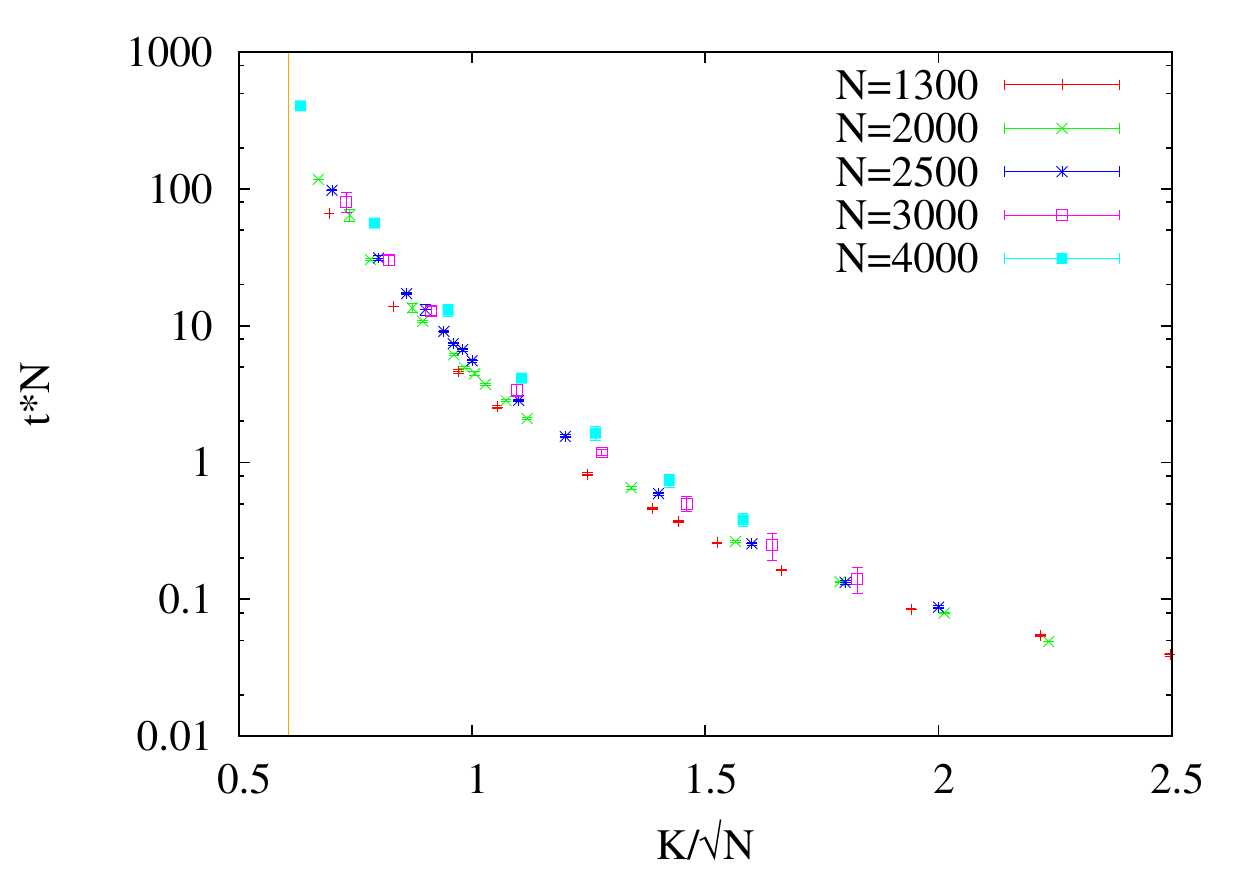}
\caption{\label{Fig:t(K)} Left: Average time $t$ needed by BayesMC for the recovery of the panted clique as a function of $K$ at the optimal temperature $\beta_{opt}(K)$ for graphs of different sizes $N$. The solid line is the best fit with a power-law function $t=\frac{a}{(K-K_{min}(N))^{\nu}}$. Averages are taken over $\sim 10^3$ samples. Right: Same data as in left. On the $x$ axis, we put $K/\sqrt{N}$ to compare data of different sizes, while the recovery times are scaled linearly with the size $N$: using such rescaled variables data at different sizes $N$ collapse in the whole analyzed region. The orange vertical line is the threshold $\frac{K}{\sqrt{N}}=\frac{1}{\sqrt{e}}$ for linear-time local algorithms.}
\end{figure}

In Fig. \ref{Fig:t(K)} we plot the value of optimal $t$, extrapolated at $\beta_{opt}(K)$, as a function of $K$ for graphs of different sizes $N$: times grow lowering $K$, and their behaviour is well described by a power-law function $t=\frac{a(N)}{(K-K_{min}(N))^{\nu}}$.
We want to check if the BayesMC working at the optimal temperature is a polynomial algorithm in the region $K=O(\sqrt{N})$. For this reason in the right part of Fig. \ref{Fig:t(K)} we plot the same data as a function of $\frac{K}{\sqrt{N}}$. Scaling the recovery times with a factor $N$, data from different sizes $N$ collapse: the numerical indications suggest that BayesMC is a polynomial algorithm for $K=O(\sqrt{N})$.
It is known that no linear-time local algorithm can find the planted clique for 
$K<K_{BP}\equiv\sqrt{\frac{N}{e}}$ \cite{deshpande2015finding,montanari2015finding}: BayesMC seems to reach this threshold.
 
\section{BP analysis}

In this section we want to gain some insights about the possible phase transitions happening in the landscape identified by the 
$\beta$-dipendent distribution in eq. (\ref{eq:betaPosterior}).
Given a posterior distribution, one can define a message passing algorithm, called Belief Propagation (BP), that should converge to the correct marginals for each node variable, if properly initialized. 
The BP equations were written and analyzed in ref. \cite{deshpande2015finding} in the case of $\beta=1$. Here we generalize
them to $\beta\neq 1$, to extract the marginal probabilities for each node from eq. (\ref{eq:betaPosterior}).
We introduce cavity messages $\psi_{i\rightarrow j}(x_i)$ that encode the probability that node $i$ takes value $x_i$,
conditioned on the absence of edge $(ij)$. 
The main assumptions of BP is that the incoming messages are probabilistically independent. In the present case the property of the graph to be dense makes single interactions sufficiently weak so that the assumption of independence 
of incoming messages is plausible at the leading order in the large size limit. 
One can then write iterative equations on the cavity messages, that read:

\begin{align*}
 \psi_{i\rightarrow j}(x_i=0)=&\frac{1}{z_{i\rightarrow j}}\left(\frac{N-K}{N}\right)^\beta\left(\frac{1}{2}\right)^{\beta(N-1)},\\
 \psi_{i\rightarrow j}(x_i=1)=&\frac{1}{z_{i\rightarrow j}}\left(\frac{K}{N}\right)^\beta\left(\frac{1}{2}\right)^{\beta(N-1)}\prod_{k\backslash j}\left[1+(2^\beta A_{ij}-1)\psi_{k\rightarrow i}(1) \right].
\end{align*}
with $z_{i\rightarrow j}$ a normalization factor ensuring that $\psi_{i\rightarrow j}(0)+\psi_{i\rightarrow j}(1)=1$.

Once the iteration of the cavity messages has reached a fixed point, marginal probabilities are obtained as:
\begin{align}
\nonumber
 \psi_{i}(x_i=0)=&\frac{1}{z_{i}}\left(\frac{N-K}{N}\right)^\beta\left(\frac{1}{2}\right)^{\beta N},\\
 \psi_{i}(x_i=1)=&\frac{1}{z_{i}}\left(\frac{K}{N}\right)^\beta\left(\frac{1}{2}\right)^{\beta N}\prod_{k}\left[1+(2^\beta A_{ij}-1)\psi_{k\rightarrow i}(1) \right].
 \label{eq:BP}
\end{align}
with $z_{i}$ a different normalization factor ensuring that $\psi_{i}(0)+\psi_{i}(1)=1$.
The \textit{Bethe free energy} associated to the reached solution is 
\begin{equation}
 f=-\frac{1}{N}\left[\sum_i\log(z_i)-\sum_{ij}\log(z_{ij})\right],
 \label{eq:f_BP}
\end{equation}
with $z_{ij}=\frac{z_i}{z_{i\rightarrow j}}$. 

At this point one can assign labels to the spins. 
A first possibility is an assignment that maximizes the posterior probability \cite{deshpande2015finding}:
$$\hat{x}^{(1)}_i=\argmax_{x=0,1}\psi_i(x)$$

However we look also at another assignment $\{\hat{x}^{(2)}\}$: supposing to know the exact
$K$, we assign $x_i^{(2)}=1$ to the first $K$ nodes ordered according to their $\psi_i(1)$.

Depending on the chosen way to assign the labels, we can define two different overlaps with the planted solution. If the assignment $\hat{x}^{(1)}$ is chosen, we define the overlap, following 
ref \cite{decelle2011PRE}, as:
\begin{equation}
    q^{(1)}=\frac{\frac{1}{N}\sum_{i=1}^N\delta_{\hat{x}_i^{(1)},v_i}-q^{(1)}_{ran}}{1-q^{(1)}_{ran}}
\label{eq:overlap1}    
\end{equation}
with $q^{(1)}_{ran}=1-\frac{K}{N}$ being the overlap of an uninformative assignment $\hat{x}_i^{(1)ran}=0\quad \forall i$,
obtained when marginals are just computed according to the prior.
On the other hand, if the assignment $\hat{x}^{(2)}$ is chosen, we define the overlap as:
\begin{equation}
    q^{(2)}=\frac{\frac{1}{N}\sum_{i=1}^N\delta_{\hat{x}_i^{(2)},v_i}-q^{(2)}_{ran}}{1-q^{(2)}_{ran}}
\label{eq:overlap2}    
\end{equation}
with $q^{(2)}_{ran}=\frac{K^2}{N^2}+\frac{(N-K)^2}{N^2}$ being the overlap of an uninformative assignment for which the $K$ nodes with $\hat{x}_i^{(1)ran}=1$, are chosen at random among the $N$ possible ones.

Let us briefly recap which is the behaviour of the BP algorithm in the $\beta=1$ case:
For $K>K_{BP}=\sqrt{N/e}$, BP always converges to a fixed-point (FP) that has a high overlap with the planted solution. For $K_d<K<K_{BP}$, BP initialized around the planted solution finds a high-overlap solution, while it fails to identify the planted solution if randomly initialized. $K_d$ should correspond to $K_d=\log_2(N)$ in the $N\to\infty$ limit \footnote{This threshold lacks of an analytical proof. Moreover $K_d$ suffers from huge finite size effects, and it has been numerically measured to be $K_d\simeq 1.3 \log_2(N)$ for $N\simeq10^4$ in ref. \cite{angelini2018parallel}.}.
One can compute the Bethe free-energy $f$ associated to the two fixed-points: for $K>K_s$ the global minimum of $f$ corresponds to the planted FP while for $K<K_s$ the FP reached from random initialization has lower free-energy. $K_s=2\log_2(N)$ in the thermodynamic limit and identifies the information-theoretical threshold below which it is impossible to recover the planted solution.
For $K<K_d$ even BP with planted initialization fails to find a high-overlap solution: the planted solution becomes locally unstable. In a statistical mechanics language $K_d$ corresponds to the spinodal point for the existence of the planted state, while $K_{BP}$ is the spinodal point for the existence of the low-overlap state.
In ref. \cite{deshpande2015finding}, the state evolution equations are rigorously derived for the evolution of the cavity messages in the $N\to\infty$ limit: 
It is proven that $\Gamma\equiv\log\left(\sqrt{N}\frac{\psi(1)}{\psi(0)}\right)$ at time $t$ is a Gaussian variable of mean $\mu_0(t)$ and variance $\sigma^2(t)$ for nodes that do not belong to the planted clique, and mean $\mu_1(t)$ and variance $\sigma^2(t)$ for nodes that do belong to the planted clique.
For $K>K_{BP}$, $\lim_{t\to\infty}\frac{\mu_1(t)-\mu_0(t)}{\sigma(t)}=\infty$ and the recovery is an easy task.
For $K<K_{BP}$ the above limit is finite, but different from 0 at least in the whole region $K=\kappa\sqrt{N}$, with $\kappa=O(1)$ \footnote{The region $K=o(\sqrt{N})$ was not studied in ref. \cite{deshpande2015finding}}: BP with random initialization does not find a high-overlap solution, however it reaches a non-trivial inference accuracy; while $q^{(1)}=0$ in this region, $q^{(2)}>0$: this is the reason why we introduced it. This is a so-called hybrid-hard phase: a phase of this type has been shown to be present in many other inference problems \cite{lesieur2017constrained,ricci2019typology}.

The transitions at $\beta=1$ can be straightforwardly generalized to $\beta\neq 1$
looking at the fixed points reached by the BP equations in eq. (\ref{eq:BP}) from random or planted initialization and to the associated free-energies. 

First of all we compute the critical temperature $T_d(K)$ above which the planted solution is not locally stable anymore:
at a given temperature $T$, we initialize the cavity messages around the planted solution, taking $\psi_i(1)=1-\delta$ for $i\in\mathcal{C}$ and $\psi_i(1)=\delta$ for $i\notin \mathcal{C}$, with $\delta$ being a random variable uniformly extracted in $\delta\in[0,10^{-4}]$.
We then run the BP iteration following eq. (\ref{eq:BP}) and check if the reached FP has high overlap with the planted solution.
For $T>T_d(K)$ the reached FP has low overlap with the planted solution.

Then we compute the critical temperature $T_{BP}(K)$ below which the low-overlap solution is not locally stable anymore:
at a given temperature $T$, we initialize the cavity messages at random, taking $\psi_i(1)=\delta$ $\forall i$  with $\delta$ being a random variable uniformly extracted in $\delta\in[0,1]$.
We then run the BP iteration following eq. (\ref{eq:BP}): For $T<T_{BP}(K)$ the reached FP always has high overlap with the planted solution.

Finally, at fixed $T$ we compare the free-energies computed following eq. ($\ref{eq:f_BP}$), associated to the FP reached from planted or random initialization. In this way we can locate the thermodynamic threshold $T_s(K)$ above which the global minimum of the Bethe free-energy is not associated to the planted solution anymore.

\begin{figure}
\centering
\includegraphics[width=0.7\textwidth]{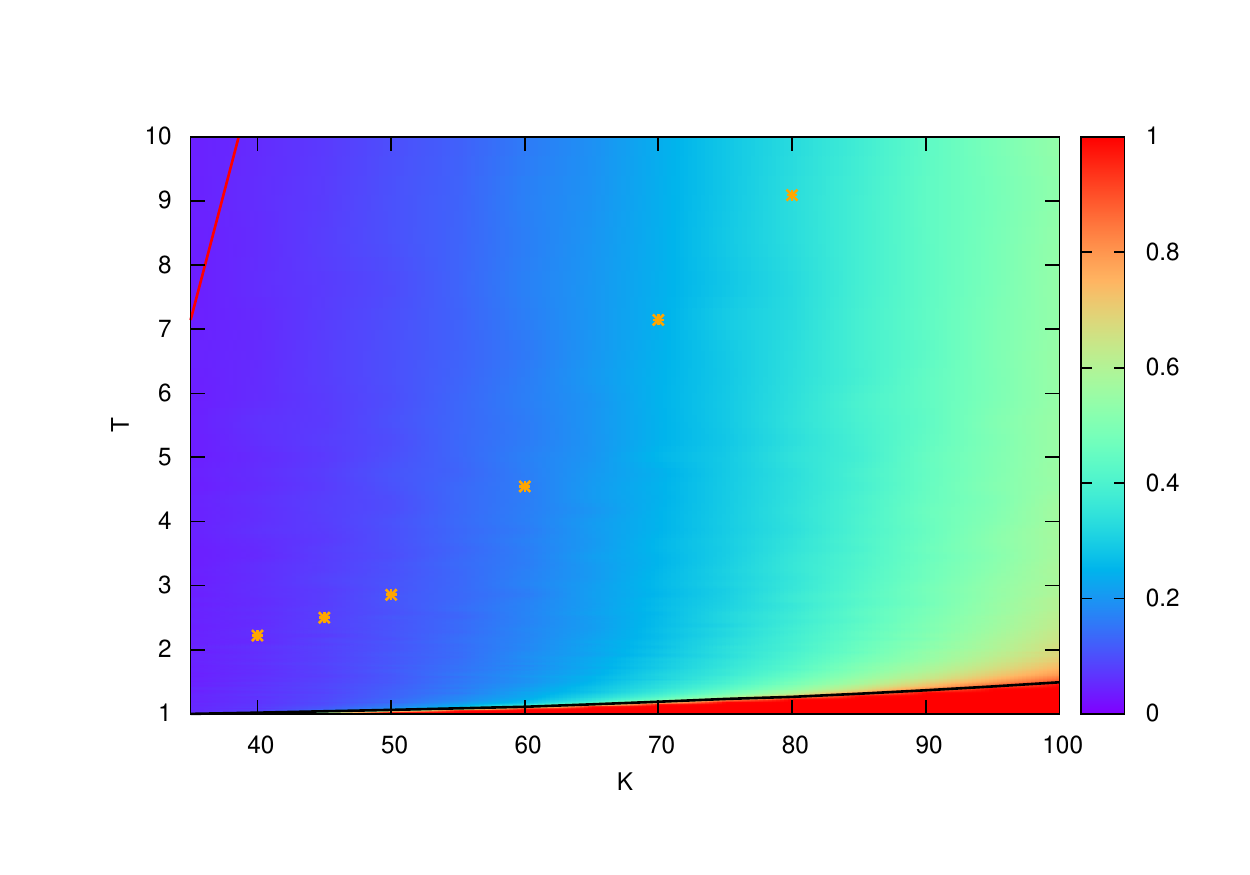}
\caption{\label{Fig:PhaseDiag} Phase diagram for $N=3000$ in the $K-T$ plane. The red line corresponds
to $T_s(K)$: for higher $T$ the recovery is impossible. $T_d(K)$ is at higher $T$, outside the range of the plot. 
The black line corresponds to $T_{BP}(K)$:
for lower $T$ the randomly initialized BP finds a high-overlap solution. 
The color scale is associated to the value of the overlap $q^{(2)}$ between the FP reached from random initialization and the planted configuration: for $T>T_{BP}(K)$ the high-overlap fixed point is missed but BP reaches a FP with small $q^{(2)}\neq0$.
The orange dots identify the optimal $T$ for the MC at fixed $K$ that always stays in the hybrid-hard region.}
\end{figure}

In Fig. \ref{Fig:PhaseDiag} we show $T_s(K)$, $T_{BP}(K)$ and the overlap $q^{(2)}$ 
for the solution found from the randomly initialized BP, together with the temperature $T_{MC}(K)\equiv \beta_{opt}^{-1}(K)$ for which the time needed by MC to find the planted solution is minimum, for systems of size $N=3000$. 
We observe that $T_{MC}(K)$ is always in the hybrid-hard phase and does not corresponds to any of the transitions extraced by the BP algorithm.
This impossibility to locate $T_{MC}(K)$ just looking at the BP fixed points and free energies can be due to two separate effects:
First of all, in physics language the BP algorithm assumes Replica Symmetry (RS). Saying it in another way, BP cannot identify spurious glassy states that can be present in the free-energy landscape. While Replica Symmetry is garanteed when one considers the posterior probability obtained having the perfect knowledge of the generative model (in this case the so-called Nishimori condition holds), and in particular Replica symmetry holds for $\beta=1$, when $\beta\neq1$ Replica Symmetry can be broken and spourious glassy states can appear, that are not detected by RS BP.
If this is the case, one should take Replica symmetry breaking into account.

Secondly, MC is a sampling algorithm that starts out-of-equilibrium. While one can analyze the stationary, equilibrium distribution, strong
out-of-equilibrium effects can affect its dynamics. The analysis of these effects is much more complicated and is a key question, also
shared with other algorithms such as (stochastic) Gradient Descent or Langevin dynamics \cite{mignacco2021stochasticity}.

\begin{figure}
\centering
\includegraphics[width=0.7\textwidth]{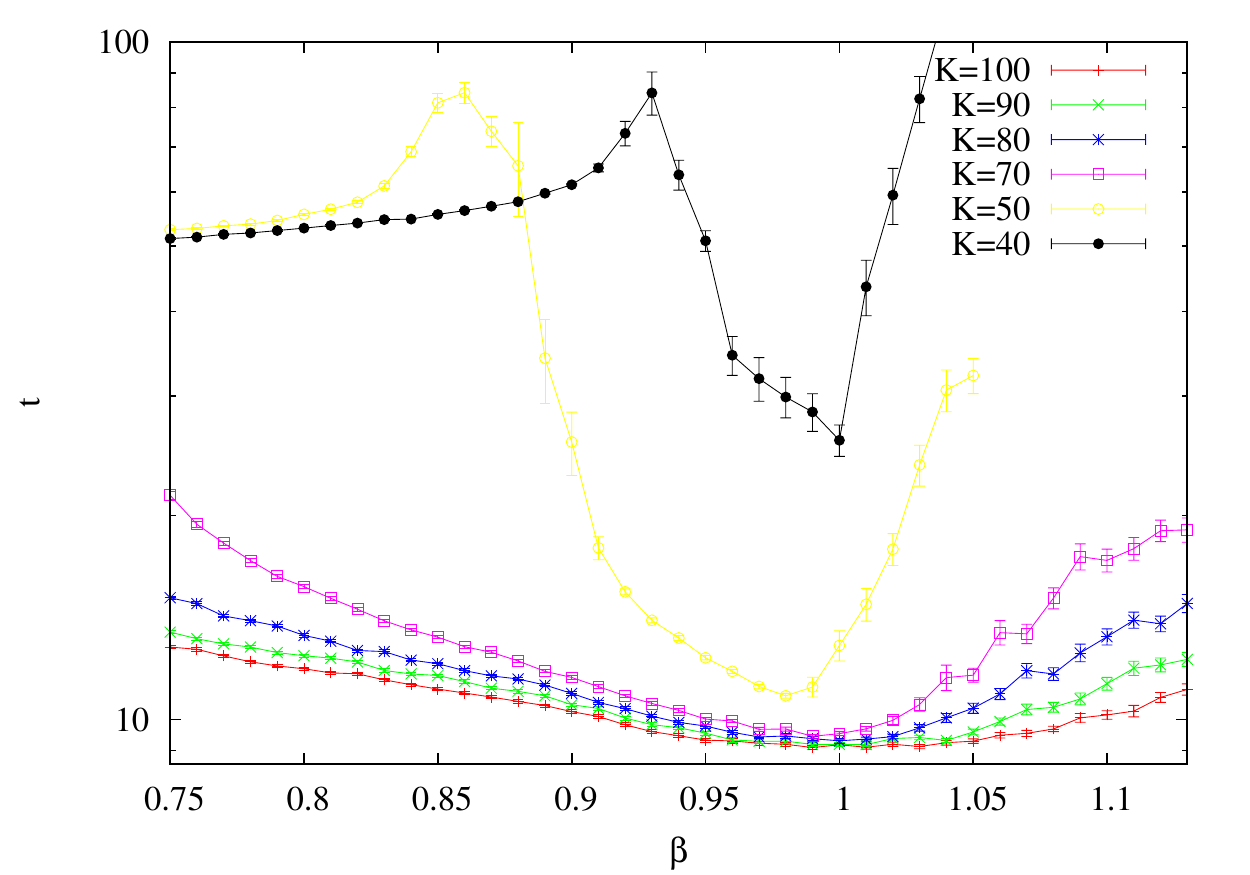}
\caption{\label{Fig:BPt} Time needed by BP algorithm to reach a fixed point starting from a random configuration for a system with $N=2000$ at different values of $K$ and inverse temperature $\beta$. The minimum is always at $\beta\simeq1$ that corresponds to the Bayes optimal case. A local maximum is present at the spinodal
point $\beta_{BP}(K)<1$ for the existence of a secondary minimum of the free-energy in addition to the planted one ($\beta_{BP}(50)\simeq 0.86$, $\beta_{BP}(40)\simeq 0.93$). For $\beta<\beta_{BP}(K)$, the BP algorithm finds a low-overlap solution, different from the planted one.}
\end{figure}

One could also wonder how BP works in the mismatched setting $\beta\neq 1$.
In Fig. \ref{Fig:BPt} we show the time $t_{BP}(\beta)$ needed by BP to reach a fixed point as a function of $\beta$: differently from MC, BP rapidly deteriorates when $\beta\neq1$, $t_{BP}(\beta)$ always showing a minimum at $\beta=1$. 
When one has the perfect knowledge of the generative model, BP is for sure the algorithm to prefer because it is able to extract the exact marginal
distribution for each node from the exact Bayes posterior: it is a so called Bayes optimal algorithm. 
However, often in real-world problems, one does not have the perfect knowledge of the model that 
generated the data. In a mismatched setting it is possible that different algorithms could outperform BP. 
Our work is showing a practical example of a mismatched setting for which MC can find the optimal solution 
to the problem in a region of the used parameters where BP cannot.

\section{Conclusions}

We showed that both Jerrum MC and BayesMC, that is a Metropolis MC based on the posterior distribution, are suboptimal in finding a planted clique: they do not succeed in polynomial time even for $K$ much higher than $K=O(\sqrt{N})$, that corresponds to the conjectured algorithmic threshold. 
The BayesMC can be generalized adding an inverse temperature parameter that takes the value $\beta=1$ 
in the original problem. We show that letting $\beta\neq1$, 
the generalized BayesMC can reach the performances of Bayes optimal algorithms such as Belief Propagation, running in a polynomial time down to $K=\sqrt{\frac{N}{e}}$. An analogous phenomenon was already shown in ref. \cite{gamarnik2019landscape} for a different class of MC algorithms, for which the assumption of a mismatched size of the clique $\overline{K}>K$ allowed the MC to reach optimal recovery. It would be nice if one could apply the same techniques of ref. \cite{gamarnik2019landscape} to analytically confirm our numerical results, and if one could link the optimal mismatched $\overline{K}$ for the MC in ref. \cite{gamarnik2019landscape} to the optimal vale of $\beta$ for the BayesMC.

Our setting of the problem as a standard statistical mechanics problem leads to a good physical intuition of what is happening: the MC works better for higher temperature, where the global minimum of the free-energy is still the planted one, but entropic effects help in moving faster through the phase space and reach it with a smaller time.

Let us emphasize that the optimal temperature for the MC to reach the planted solution is on a region where standard Bayes optimal algorithms such as Belief Propagation are stuck in a low-overlap solution: this is a clear example that when we do not have the perfect knowledge of the parameters of the model, the celebrated BP algorithm can work worse than MC.

We do not have a theoretical way to identify the optimal temperature for MC to work, and we only extracted it numerically. Taking inspiration from recent literature, the failure of MC for $T\simeq 1$ could be due to the presence of spurious glassy states that trap the dynamics \cite{GlassyNatureAntenucci}. Glassy states are not identified by BP algorithm that is based on a Replica-Symmetric assumption. This Replica-Symmetry holds at $\beta=1$ (it corresponds to the so called Nishimori line \cite{zdeborova2016statistical}), but it is broken as long as $\beta\neq 1$. 
In some recent papers \cite{mannelli2019afraid, sarao2020complex} it is shown how the performances of gradient descent (GD) based algorithms in some planted problems can be affected by spurious local minima that trap the dynamics. However, at high enough signal to noise ratio, the spurious minima at higher energy become saddles with a negative direction towards the planted state. For this reason, GD algorithms can reach the planted state, surfing on these high energy saddles that lead them directly on the planted state, even in the presence of stable local minima uncorrelated with the signal at lower energy. It could be worth looking whether a similar phenomenon happens in this case, being higher energy states more probably visited at higher temperatures.
In ref. \cite{venturi2018spurious}, the authors analyze the topology of the loss function of a neural network, showing as, although spurious valleys may exist, they are avoided with high probability on over-parametrized models. 

Usually a simple way to identify the \textit{dynamical} threshold $T_d$ for the existence of glassy states is the following: one plants a glassy state at $T_d$, then run BP initialized inside the state and see if the state remains stable under iteration or one eventually goes back to the paramagnetic fixed point. Unfortunately we cannot use this method in the planted clique problem, because we have seen how the paramagnetic phase is modified by the introduction of the planted clique, acquiring a small positive overlap with the planted solution. To identify the temperature for the birth of glassy states one should indeed run the 1-step Replica Symmetry Breaking  equations \cite{1RSB_Mezard} in presence of the planted state. We left this as future work.

Finally we would point out that the present analysis on the role of temperature in the MC behaviour could help to shed light into the behaviour of more complex MC based algorithms such as Parallel Tempering (PT) and to explain why PT numerically finds solutions to the planted clique problem down to the information-theoretical threshold as shown in ref. \cite{angelini2018parallel}. We are currently investigating this point.

We thank David Gamarnik and Federico Ricci-Tersenghi for very interesting discussions.
This research has been supported by the European Research Council under the European Unions Horizon 2020 research and innovation programme (grant No.~694925 -- Lotglassy, G. Parisi).


\bibliography{biblio}
\clearpage
\appendix
\section{Numerical analysis of the Jerrum algorithm}
\begin{figure}
\centering
\includegraphics[width=0.7\textwidth]{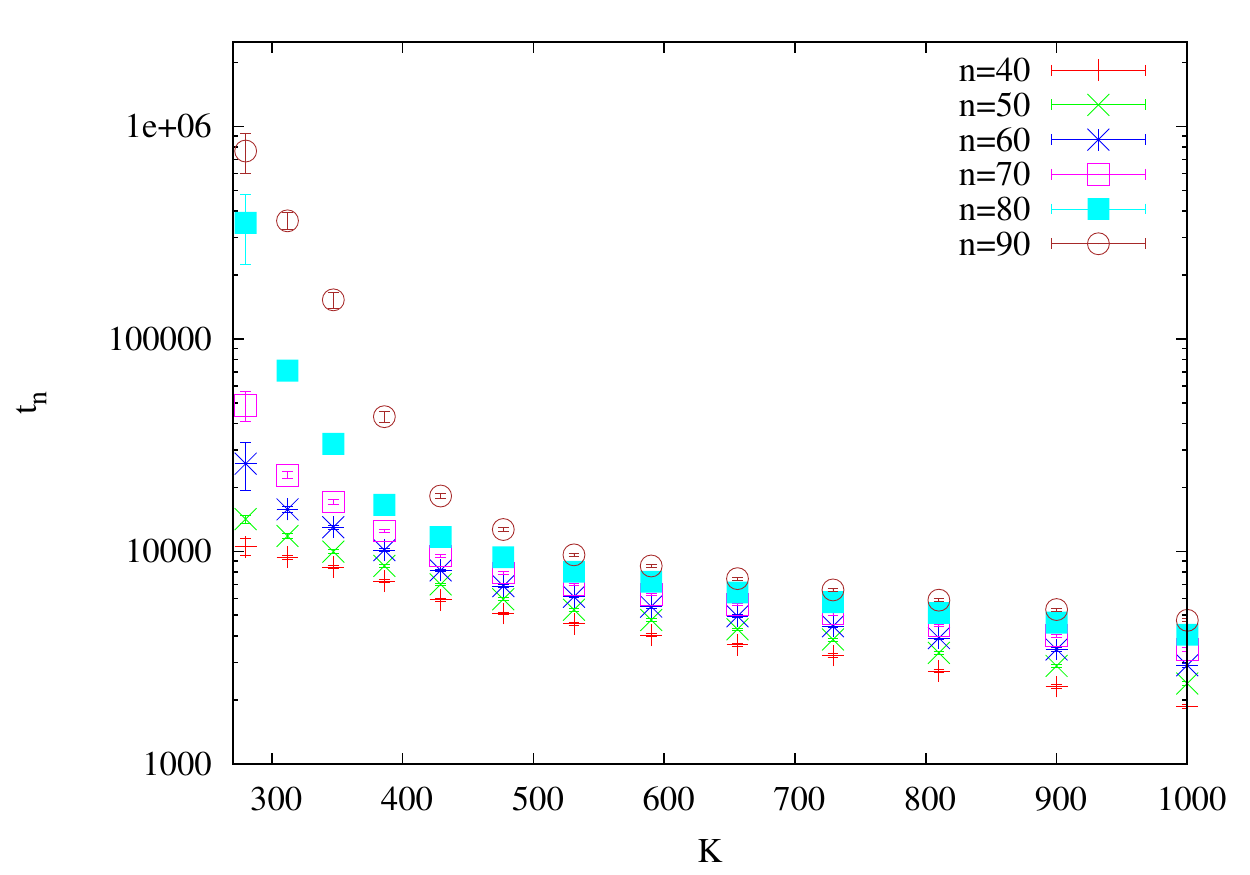}
\caption{\label{Fig:JerrumPercentile} Average time needed by the Jerrum algorithm to reach the planted solution for a system with $N=3000$ at different values of $K$: 
times $t_n$ are averaged over the fastest $n\%$ of the samples, curves are shown for different $n$. For $K\lesssim500$ the average is dominated by slowest samples and curves at different values of $n$ grow differently decreasing $K$.}
\end{figure}

\begin{figure}
\centering
\includegraphics[width=0.48\textwidth]{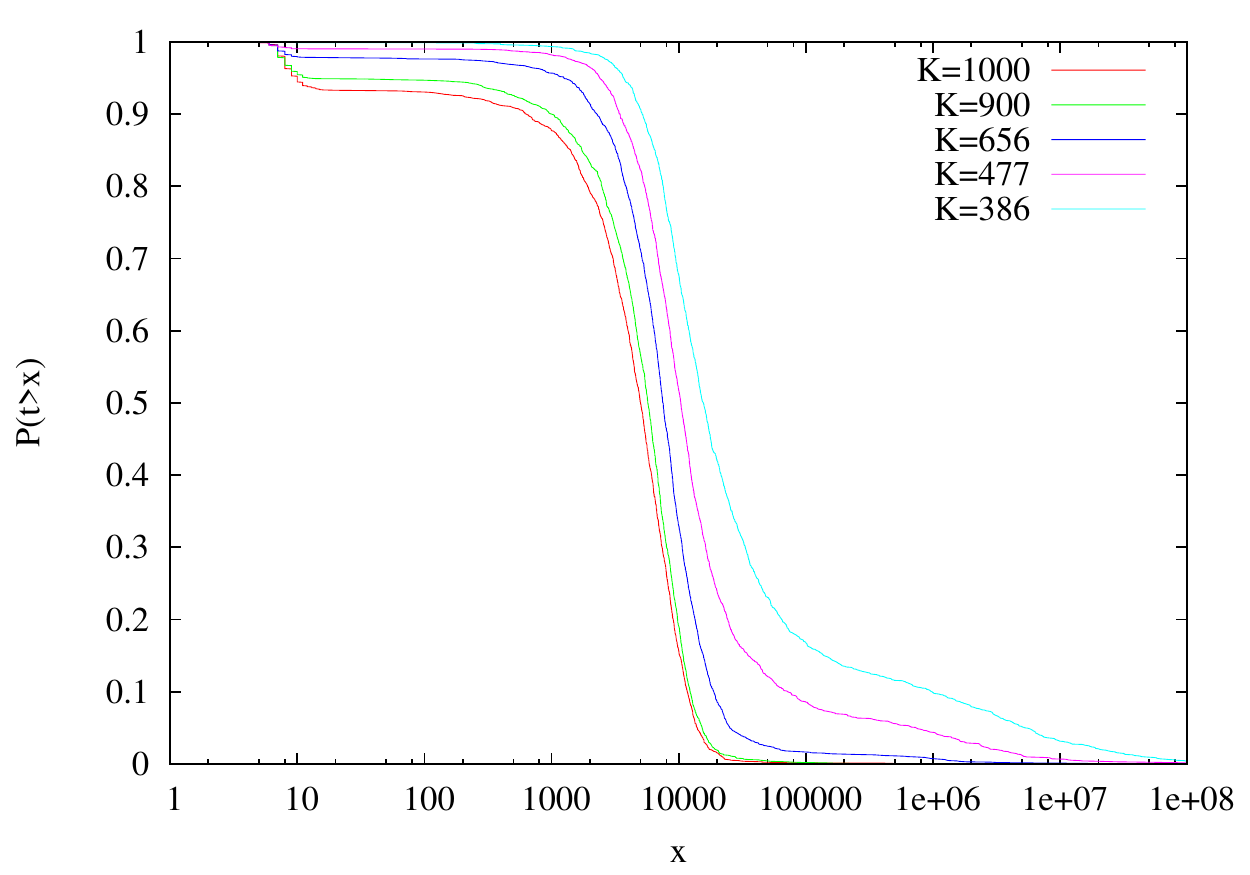}
\includegraphics[width=0.48\textwidth]{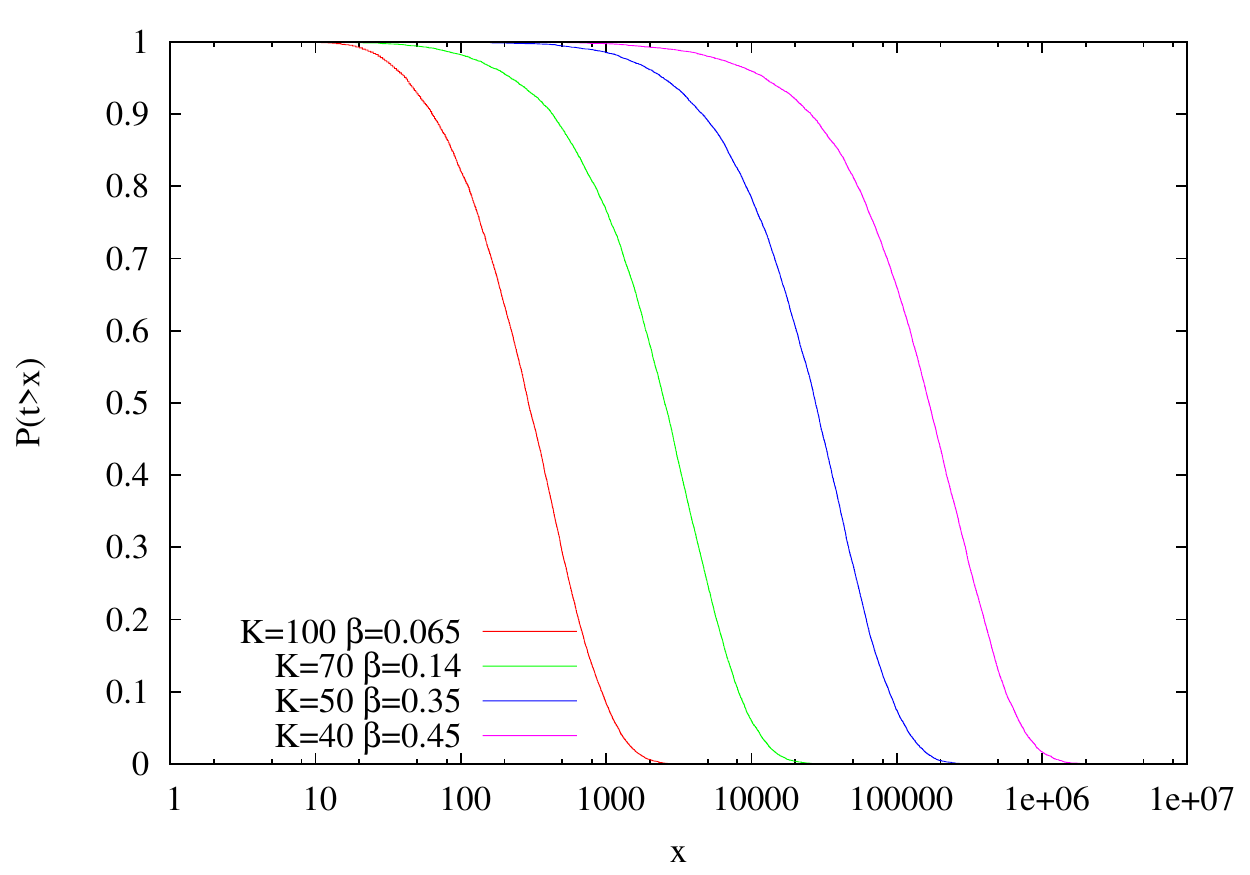}
\caption{\label{Fig:cumulJerrum} Left: Cumulative probability $P(t>x)$ that the time $t$ needed by the Jerrum algorithm to reach the planted solution for a given system with $N=3000$ would be greater than $x$, at different values of $K$:  while the median smoothly increases decreasing $K$, alongside the distributions develop heavy tails at very large times. When this happens, the averaged time is no longer concentrated around the typical case. Right: Cumulative probability $P(t>x)$ for the BayesMC for systems with $N=3000$ at the optimal inverse temperature $\beta_{opt}(K)$. The heavy tails phenomenon is not present in this case.}
\end{figure}

We present here a numerical analysis of the Jerrum Monte-Carlo algorithm \cite{Jerrum92} 
for the recovery of a planted clique of size $K$ inside a graph of size $N$. 
The acceptance probability for a move for Jerrum algorithm is defined in the main text in eq. (\ref{eq:Jerrum}). The natural quantity to measure to describe the algorithmic performances would be the average time needed to find the planted clique as a function of $K$. However we notice that, decreasing $K$, the average time over a large number of samples is dominated by worst cases: in Fig. \ref{Fig:JerrumPercentile} we show the average over the fastest $n\%$ samples, for different values of $n$: the growing rate, decreasing $K$, strongly depend on $n$. The same phenomenon is evident when  plotting the cumulative probability $P(t>x)$ that the time to solution $t$ for a given sample is greater than a certain value $x$, as shown in Fig. \ref{Fig:cumulJerrum}: while the median smoothly increases decreasing $K$, alongside the distributions develop heavy tails at very large times. When  this  happens,  the  time averaged over all the samples is  no  longer  concentrated  around  the typical case but is 
dominated by rare slow events. This situation of hardness driven by rare events, is quite common
in the study of disordered systems at finite sizes \cite{fernandez2013temperature}.

\begin{figure}
\centering
\includegraphics[width=0.7\textwidth]{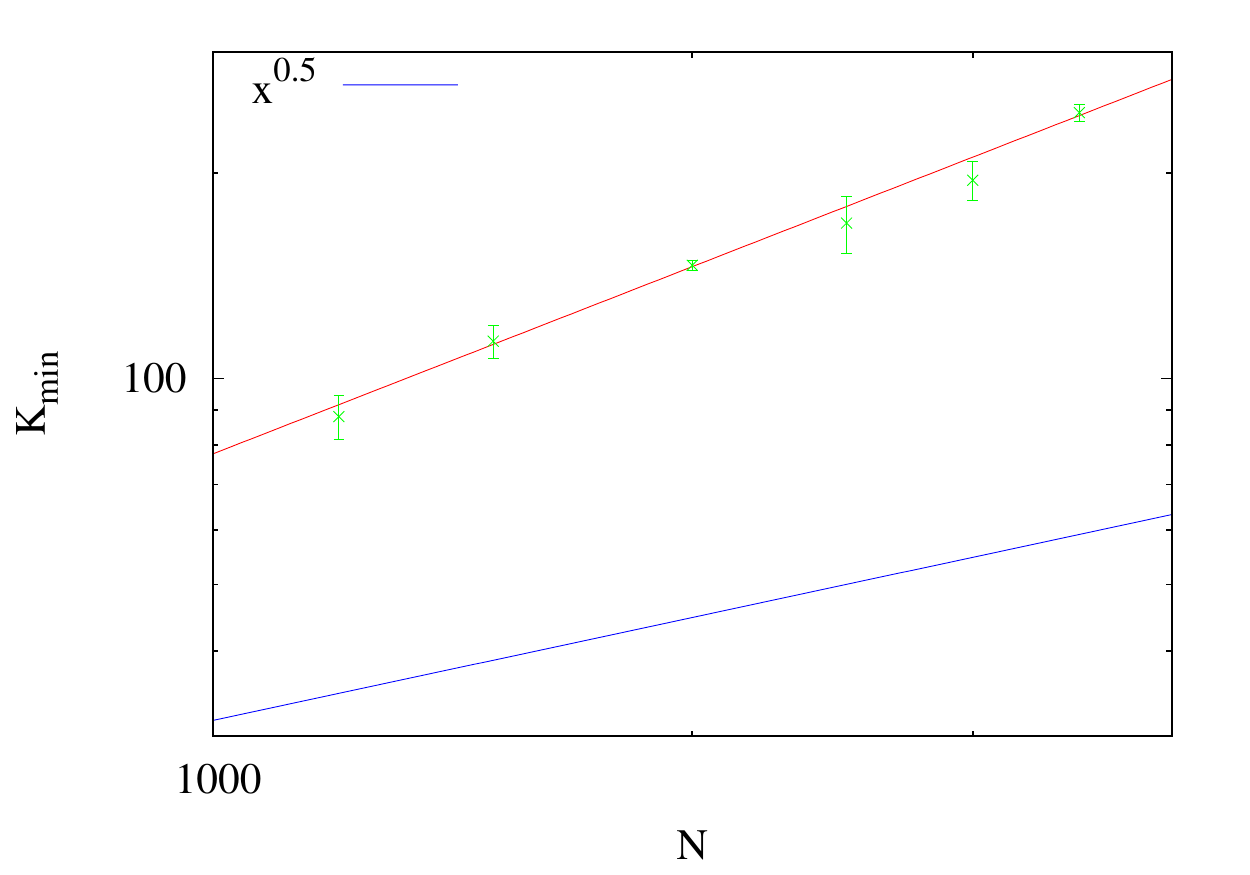}
\caption{\label{Fig:K_min_Jerrum} 
Minimum clique size $K_{min}$ that can be found by the Jerrum algorithm as a function of the graph size $N$. The red solid line is the best fit of the form in eq. (\ref{eq:FitKmin}) while the blue line is the analytical lower bound computed by Jerrum \cite{Jerrum92}.
}
\end{figure}

The heavy tails make the quantification of average times an impossible task. For this reason
we choose to look at the average time over the fastest $50\%$ of the samples, $t_{50}$, that is the quantity shown in Fig. \ref{Fig:K_min_Jerrum}. 
With this choice we disregard the heavy tail phenomenon that could be strongly related to finite size effects. In practice we are underestimating the minimum clique size $K_{min}$ that could be found by the Jerrum algorithm. However, all we want to show is that the scaling of such threshold is higher than $O(\sqrt{N})$. We are aware that it is very hard to deduce the exact form of the scaling of the threshold just by numerical experiments, specially for a problem like the planted clique with very large finite size effects and it is beyond the scope of this paper. 
For this reason, we focus on $t_{50}$, that is not the perfect choice but it leads to a lower bound on the estimation of $K_{min}$.
\begin{figure}
\centering
\includegraphics[width=0.7\textwidth]{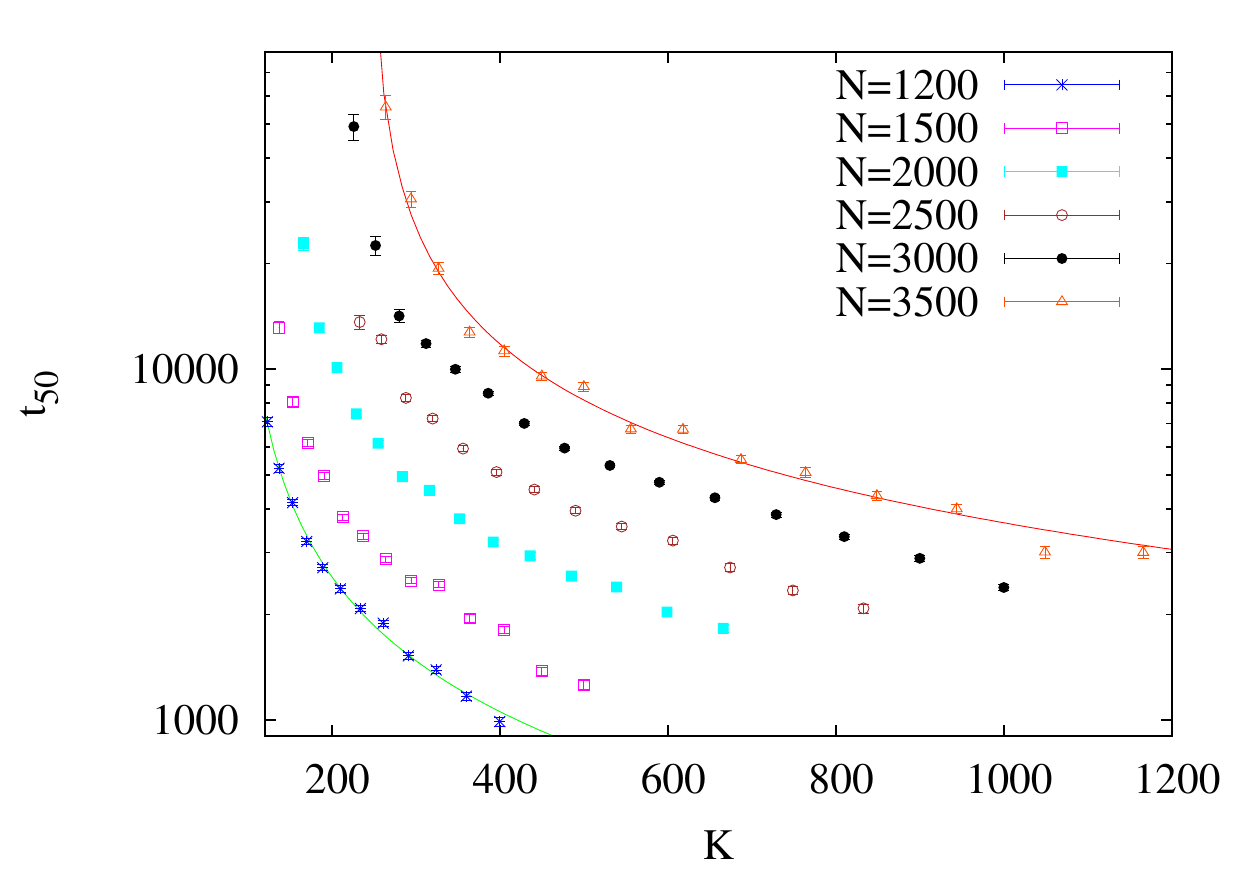}
\caption{\label{Fig:t_vs_K_Jerrum} Time to solution $t_{50}$ averaged over the fastest $50\%$ of the samples as a function of the size of the clique for different values of $N$. Solid lines are the best fit
of the form in eq. (\ref{eq:fit}).}
\end{figure}

$t_{50}$ diverges lowering $K$. We fit its behaviour with a function of the type:
\begin{equation}
t_{50}(K)=\frac{a}{(K-K_{min})^{\nu}}.
\label{eq:fit}
\end{equation}
with $a$ and $K_{min}$ depending on $N$.
In this way we can extrapolate the threshold $K_{min}(N)$ that we plot as a function of $N$ in Fig. \ref{Fig:K_min_Jerrum}.
Assuming a behaviour of the type:
\begin{equation}
    K_{min}(N)=b N^{\alpha},
\label{eq:FitKmin}
\end{equation}
we obtain an estimate $\alpha=0.91(4)$ that is much higher than the lower bound computed analytically by Jerrum $\alpha_L=0.5$.
The Jerrum MC is thus suboptimal, as already conjectured in ref. \cite{gamarnik2019landscape}. As already said, numerical estimation of the threshold exponent $\alpha$ is affected by strong finite size effects. The natural methodological framework for its analytical computation would be the one already used in ref. \cite{gamarnik2019landscape} for a different class of MC algorithms.

\end{document}